%Paper: astro-ph/9305003
%From: p165bie@mpifr-bonn.mpg.de (Peter Biermann)
%Date: Thu, 6 May 93 16:31:24 +0200

%%arXiv: hacked to compile, 20Feb2002

\input aa.cmm
\hyphenation{pro-pa-gat-ing}
\def\simle{\lower 2pt \hbox {$\buildrel < \over {\scriptstyle \sim }$}}
\def\simge{\lower 2pt \hbox {$\buildrel > \over {\scriptstyle \sim }$}}
\MAINTITLE={Cosmic Rays}
\SUBTITLE={II. Evidence for a magnetic rotator Wolf Rayet star origin}
\AUTHOR={\ts Peter L. Biermann@1 and \ts Joseph P. Cassinelli@2}
%arXiv: \SENDOFF={\ts Peter L. Biermann}
\INSTITUTE={@1 Max Planck Institut f\"ur Radioastronomie, Auf dem H\"ugel 69,
D-5300 Bonn 1, Germany @2 Department for Astronomy, University of
Wisconsin, Madison, Wisconsin 53706, USA}
%arXiv: \RECDATE={   }
%arXiv: \ACCDATE={   }
%arXiv: changed \SUMMARY to \ABSTRACT
\ABSTRACT={ Based on 1) a conjecture about the mean free path for particle
scattering in perpendicular shock geometries, and 2) a model for Wolf Rayet star
winds, we argue that explosions of Wolf Rayet stars can lead through
diffusive particle acceleration to particle energies up to $3 \; 10^{9} \; \rm
GeV$.  As a test we first demonstrate that the magnetic fields implied by this
argument are compatible with the dynamo limit applied to the convective
interior
of the main sequence stars which are predecessors of Wolf Rayet stars; second
we
show that this implies the same strength of the magnetic fields as is suggested
by the magnetically driven wind theory. Third, we use data from radio
supernovae
to check the spectrum, luminosity and time dependence which suggest that the
magnetic fields again have to be as high as suggested by the magnetically
driven wind theory.  This constitutes evidence for the magnetic field strengths
required to accelerate particles to $3 \, 10^9$ GeV.  Fourth, we demonstrate
that within our picture the nonthermal radio emission from OB and Wolf Rayet
stars can reproduce the proper radio spectra, time variability and radio
luminosities. Fifth, the comparison of Wolf Rayet stars and radio supernovae
suggests that electron injection into the acceleration process is a step
function of efficiency with the shock speed as the independent parameter; the
critical speed appears to be that speed at which the thermal downstream
electrons become relativistic. We make detailed predictions on the temporal and
spectral behaviour of the nonthermal radio emission of OB and Wolf Rayet stars
that will allow further checks our model.}

\KEYWORDS={Plasma Physics -- Shockwaves -- Supernovae -- Wolf-Rayet stars --
Cosmic Rays}

\THESAURUS={02.16.1, 02.19.1, 08.19.4, 08.23.2, 09.03.2}

\maketitle
\titlea {Introduction}
The origin of cosmic rays above particle energies of about $5 \; 10^{6} \; \rm
GeV$ is still not understood.  The main proposal that has been introduced by
Jokipii and Morfill (1987) argues that there is a terminal shock of the
galactic
wind which accelerates particles much like the terminal shock of the solar
wind,
and that then these particles diffuse against the wind "down to us".  In this
picture it is unclear how the intensity of the cosmic ray flux at the matching
point of $5 \; 10^{6} \; \rm GeV$ can fit to the lower energy particles which
are believed to arise from direct explosions into the interstellar medium
(Lagage and Cesarsky 1983) and from explosions of massive stars into their
stellar wind cavity (V\"olk and Biermann 1988, Silberberg et al. 1990). In the
latter theory one readily reaches particle energies right up to the
knee of the cosmic ray spectrum at $5 \; 10^{6} \; \rm GeV$.
\par
Motivated by recent developments in magnetic rotator theory, summarized by
Cassinelli (1991, 1993), that account for the strong winds of Wolf Rayet stars,
but which require much stronger magnetic fields than had been assumed by V\"olk
and Biermann (1988), we argue in this paper that explosions of Wolf Rayet stars
and similar massive stars with strong winds may well account for the
dominant cosmic ray component above particle energies of about $10^4$ GeV all
the
way to about $3 \; 10^{9} \; \rm GeV$.
\par
At energies above this presumably the extragalactic component takes over
(Biermann 1992, Rachen and Biermann 1992, 1993, paper UHE CR I, Rachen 1992,
Rachen et al. 1993, paper UHE CR II).
\par
Here we propose to present the case in several steps:  First, we
demonstrate that the dynamo mechanism applied to the interior of massive stars
-
which later get exposed by mass loss - readily can account for magnetic field
of
substantial magnitude. Then we argue how this leads to the high particle
energies; we discuss the spectrum of the cosmic rays between $10^{4} \;
\rm
GeV$
and $3 \; 10^{9} \;
\rm GeV$ in paper CR I (Biermann 1993); the consequences
for
the chemical abundances and spectrum of the cosmic rays have been checked
against air shower data by Stanev et al. (1993, paper CR IV).
\par
Second, we check that the notion that massive stars have strong
magnetic fields in their winds, as suggested by Hartmann and Cassinelli(1981)
and Cassinelli (1982) and implied by our model for the origin of energetic
Cosmic Rays (papers CR I and CR IV), is consistent with other properties of WR
stars, and by implication, also radio supernovae and OB stars.  As a test we
derive the properties of the expected nonthermal radio emission and show that
this is in agreement with the available data; further radio observations may
provide an additional detailed test for our concept.
\par
The paper is structured as follows:  In section 2 we briefly summarize the
basic properties of a fast stellar wind with a strong Parker type magnetic
field.  In sections 3 to 6 we derive the spectrum of energetic particles
accelerated in shocks that traverse such winds, fully analoguous to paper CR
I.  In section 7 we use the dynamo limits to argue that these high magnetic
fields are indeed plausible.  In section 8 we derive the implications for the
momentum of winds, and show that these winds are quite likely driven by the
magnetic field.  In section 9 we derive the expected nonthermal radio emission,
and also discuss optical thickness effects.  In sections 10 and 11 we use data
on radiosupernovae to derive numerical estimates for the magnetic field
strength in Wolf Rayet star winds independently.  In section 12 we then discuss
the nonthermal radio emission from Wolf Rayet stars, and in section 13 from OB
stars.  We finally describe in section 14 further possible tests using detailed
radio monitoring of Wolf Rayet and OB stars, and conclude with a summary in
section 15.

\titlea {Wolf Rayet star winds}

The winds of Wolf Rayet stars are well established by optical, infrared and
radio
observations.  They show a typical mass loss rate of about $10^{-5} \; \rm
M_{\odot} \; year^{-1}$, a typical wind velocity of $0.01 \, c$ and thermal
radio
emission of several mJy at $5 \;\rm GHz$ at a typical distance of $1 \; \rm
kpc$.  These numbers lead to a reference density in the wind of $7.3 \; 10^6 \;
\rm cm^{-3}$ at a radius of $10^{14} \; \rm cm$ at $5$ GHz.  This implies a
radius where optical thickness unity for free-free emission - the thermal radio
emission - of $1.4 \; 10^{14} \; \rm cm$.
\par
We assume that these winds have a strong magnetic field in the configuration of
a Parker spiral with a magnetic field that varies as $1/r$ near the equator and
is predominantly azimuthal there and as $1/r^2$ near the pole where it is
mostly
radial:

$$B_r \;=\;B_o \; (r_o/r)^2 \eqno\autnum $$

$$B_{\phi}\;=\;B_o \; (r_o/r) \; sin {\theta} \eqno\autnum $$
where $\theta$ is the colatitude, with

$$r_o \;=\; V_W/\Omega_s \eqno\autnum $$
where $V_W$ and $\Omega_s$ are the wind velocity and the angular rotation rate
of
the star itself.  Obviously, if $r_o < r_s$ then we have to
replace $r_o$ with the stellar radius $r_s$.  The magnetic
field inside $r_o$ if $r_o > r_s$ is predominantly radial
and decreases with $1/r^2$.
\par
The Alfv\'en velocity with respect to the generally stronger tangential
magnetic
field is then given by

$$v_{A \phi}\;=\;2.1 \;10^8 \,{\rm cm \,sec^{-1}}\, B_{0.5}\, ({{\dot {M}}_{-5}
\over V_{W,-2}})^{-1/2} , \eqno\autnum $$
which is independent of radius in the region where the wind velocity is
constant.  Here the mass loss $\dot{M}$ is in units of $10^{-5} \; \rm
M_{
\odot}
\; year^{-1}$, and the wind velocity in units of $0.01 \, c$.  From paper CR I
we also use the estimate that the magnetic field is $3$ Gauss at the reference
radius of $10^{14}$ cm, and so we use this value as reference, since it is the
consequences of this argument which we propose to test.  It is interesting to
note that this Alfv\'en velocity is rather close to the wind velocity, and we
will argue below that there is a physical reason for this.
\par
We propose to interpret the nonthermal radio emission both from normal Wolf
Rayet stars as well as from supernova explosions into such winds as being
caused
by particle acceleration and synchrotron emission; hence we have to consider
the
properties of shockwaves in Wolf Rayet stars.  It follows that a shockwave
propagating through such a wind encounters for most of its area a magnetic
field
which is perpendicular to the shock direction, a configuration in which
classical Fermi acceleration is known not to work (Drury 1983).

\titlea {The Cosmic Ray particle energies}

In the following four sections we reformulate the theory of particle
acceleration for perpendicular shocks introduced by Biermann (1993, paper CR I)
and apply it in detail to shocks in winds of massive single stars.
\par
In paper CR I we have introduced the conjecture that the scattering of
energetic particles in a perpendicular configuration is diffusive and that the
scattering coefficient $\kappa_{1,2}$ is composed of the radial scale of the
region and the velocity difference across the shock, and is independent of
energy:  Generalizing now for arbitrary wind speed and arbitrary shock strength
we obtain for the thickness of the shocked layer

$$\Delta r/r\;=\;{U_2 \over U_1}\;{U_1 \over {V_W+U_1}}. \eqno\autnum $$
This reduction of the thickness of the
layer for a finite wind velocity is due to the fact that the material which is
snowplowed together is not all gas between zero radius and the current radius
$r$, but between zero radius and $r (1-{V_W / (V_W+U_1)})$, since the gas keeps
moving while the shock moves out towards $r$.  This then leads to

$$\kappa_{rr,1}\;=\;{1 \over 3}\;U_1\;r\;(1 -{U_2 \over U_1})\;/
(1+V_W/U_1) \eqno\autnum $$
and

$$\kappa_{rr,2}\;=\;{1 \over 3}\;U_2\;r\;(1 -{U_2 \over U_1})\;/
(1+V_W/U_1). \eqno\autnum $$
This then results in the sum of the residence times on the two
sides of the shock front, of:

$${4\kappa_{rr,1} \over U_1 c}\;+\;{4\kappa_{rr,2} \over U_2 c}\;=\;
{8 \over 3}{r \over c} (1-{U_2 \over U_1})\;/ (1+V_W/U_1). \eqno\autnum $$

The maximum particle energy corrected for the wind
velocity then is given by a spatial limit

$$E_{max}\;=\; Z e r B_1 \;/ (1+V_W/U_1). \eqno\autnum $$
\par
We emphasize that in the Parker spiral regime of the wind the product of the
magnetic field strength and the radius is a constant, i.e. does not vary with
radius, and so we can use any radius well outside $r_o$ as reference.  Now in
order that we can reach particle energies of $3 \; 10^{9} \; \rm GeV$ we
then require (see paper CR I and eq. 9) that the product of $B$ and $r$ is
$10^{16} \; \rm Gauss \; cm$ for protons, or $3\;10^{14}\;\rm Gauss \; cm$
for iron nuclei.  We will use in the following the assumption that only iron
nuclei (an independent observational check using air shower data on the
chemical
abundances of high energy cosmic rays is done in paper CR IV) are the highest
energy particles, and so use $B=3\;\rm Gauss$ at
$r=10^{14}\;\rm cm$ as reference.

\titlea {Particle Drifts}

Consider particles which are either upstream of the shock, or
downstream; as long as the gyrocenter is upstream we will consider the
particle to be there, i.e. upstream, and similarly downstream. Following paper
CR I we assume that drifts are due to both the radial gradient of the
magnetic field and the increased curvature due to the turbulence.  This
increases the $\theta$-drift to

$$V_{d,\theta}\;=\;f_d\, c \,r_g/r
\eqno\autnum$$
from the lower level based on pure gradient drift, where

$$f_d\;=\;{1 \over 3}\,(1+{1 \over 6}\,{{r U_1} \over {\kappa_{rr,1}}}\,{U_1
\over U_2}\,(1-{U_2 \over U_1})).   \eqno\autnum$$
It is easily verified that $f_d=1$ for strong shocks and negligible wind
velocity.  We take here the curvature length scales to be the same on both
sides
of the shock, since the curvature is induced by the thickness of the shock
region.
\par
The energy gain associated with the $\theta$-drift is given by the
product of the drift velocity, the residence time, and the electric field.
Upstream and downstream together this energy gain is given by

$$\Delta E/E\;=\;4\,{U_1 \over c}\,{{\kappa_{rr,1}} \over {r U_1}}\,f_d\,
(1+{U_2 \over U_1}). \eqno\autnum $$

\titlea {The energy gain of particles}

Let us consider then one full cycle of a particle remaining near the shock and
cycling back and forth from upstream to downstream and back. The energy gain
just due to the Lorentz
transformations in one cycle can then be written as

$${\Delta E \over E}_{LT}\;=\;{4 \over 3} {U_1 \over c} (1 -{U_2 \over U_1}).
 \eqno\autnum $$
Adding the energy gain due to drifts we obtain

$${\Delta E \over E}\;=\;{4 \over3} {U_1 \over c} (1 -{U_2 \over U_1})
x \eqno\autnum $$
where

$$x\;=\;1+3\,{{\kappa_{rr,1}} \over {r U_1}}\,f_d\,
(1+{U_2 \over U_1}) / (1-{U_2 \over U_1}) , \eqno\autnum $$
which is $9/4$ for negligible wind speeds and a strong shock when $U_1 / U_2
\;=\;4$; on the other hand, for $V_W / U_1=1$ we have $x=2.042$ and for the
limiting case of large wind speeds compared to shock speeds $x=1.833$.

\titlea {Expansion and injection history}

Consider how long it takes a particle to reach a certain energy:

$${dt \over dE}\;=\;\lbrace 8 \;{\kappa_{rr,1} \over U_1 c} \rbrace /
\lbrace
{4
\over 3} {U_1 \over c} (1-{U_2 \over U_1}) x E \rbrace. \eqno\autnum $$
Here we have used that $\kappa_{rr,1} / U_1 \;=\; \kappa_{rr,2} / U_2$.
\par
Since we have

$$r\;=\;U_1 \,(1+{V_W\over U_1}) \, t \eqno\autnum $$
this leads to

$${dt \over t}\;=\;{dE \over E} {{3 U_1} \over {U_1 - U_2}} {2 \over x}
{{\kappa_{rr,1}} \over {r U_1}} \, (1+{V_W\over U_1}) \eqno\autnum $$
and so to a dependence of

$$t(E)\;=\;t_o \; ({E \over E_o})^{\beta} \eqno\autnum $$
with

$$\beta\;=\;{{{3 U_1} \over {U_1 - U_2}} {2 \over x} {{\kappa_{rr,1}} \over
{r U_1}}} \, (1+{V_W\over U_1}). \eqno\autnum $$
\par
Particles that were injected some time $t$ ago
were injected at a different rate, say, proportional to $r^b$. Also, in
$d$-dimensional space, particles have
$r^d$ more space available to them than when they were injected.  This
then leads to a combined correction factor for the abundance of

$$({E \over E_o})^{- (b+d) \beta}. \eqno\autnum $$
The combined effect is a spectral change by

$${{3 U_1} \over {U_1 - U_2}} {2 \over x} (d+b) {{\kappa_{rr,1}} \over {r U_1}}
\, (1+{V_W\over U_1}).  \eqno\autnum $$
\par
Hence the total spectral difference, as compared with the planeparallel case,
is
given by

$${{3 U_1} \over {U_1 - U_2}}\;\lbrace {U_2 \over U_1} ({1
\over x} -1) \;+\;
{2
\over x} (b+d) {{\kappa_{rr,1}} \over {r U_1}} \, (1+{V_W\over U_1}) \rbrace.
\eqno\autnum $$
Here the sign convention is such (see paper CR I) that for this expression
positive the spectrum is steeper.  For a wind we have $b+d \,=\, 1$.  We note
that

$$\kappa_{rr,1}(1+{V_W\over U_1}) / {r U_1}   \eqno\autnum $$
is now independent of the wind speed, and
the only effect of the wind which remains is through $x$. For the sequence of
$V_W/U_1\,=\,0.,1.0, \,{\rm and}\, >>1$ we thus obtain particle spectral index
differences, in addition to the index of $7/3$, of
$0.0,\,0.136,\,0.303$, corresponding to Synchrotron emission
spectral index of an electron population with the same spectrum, of
$0.667,\,0.735,\,0.818$.  We will use the first two cases as examples
below, and since the work of Owocki et al. (1988) suggests that typical shocks
in winds have a velocity in the wind frame similar to the wind velocity in the
observers frame itself, which implies that, in the simplified picture here,
only
spectral indices for the synchrotron emission between $0.667$ and $0.735$ are
relevant, with an extreme range of spectral indices up to $0.818$ for strong
shocks.  Obviously, for weaker shocks with $U_1/U_2 \,<\,4$ the spectrum can be
steeper, e.g. for $U_1/U_2 \,=\,3.5$ we obtain an optically thin spectral index
for the synchrotron emission of $0.734$ for $V_W\, \ll \,U_1$ and $0.815$ for
$V_W/U_1 \, = \, 1$.

\titlea {The strength of the magnetic field}

First we propose to demonstrate that high magnetic fields can be generated
inside massive stars so that later, when these interiors get exposed to become
the surface of Wolf Rayet stars these high magnetic fields can drive the wind.
The dynamo mechanism acts in the turbulent zone in the interior of rotating
massive stars and, given a seed field, increases the magnetic field strength up
to the limit where the Coriolis force equals the magnetic stresses (Ruzmaikin
et
al. 1988, Gilbert and Childress 1990, Gilbert 1991).  The strength of the
magnetic field can be estimated from the condition, that the magnetic
torque is limited by the Coriolis forces over the size of the convective
region.  This condition can be written as

$$B\;=\;(\Omega \,\rho_c \,R_{cc} \,v_t \,f_1)^{1/2}  , \eqno\autnum $$
where $\Omega$ is the rotation rate of the star, $
\rho_c$ is the average
density
of the convection zone, $R_{cc}$ is the radius of the convective core, $v_t$ is
the characteristic turbulent velocity, and $f_1$ is a correction factor of
order
unity in order to allow for a) structural variations ignored here, and also for
b) the fact that the dominant length scale is likely to be smaller than the
radius of the convective region.  The turbulent velocity is estimated from the
condition that turbulent convection transports all the luminosity $L$:

$$L\;=\;4 \pi \,R_{cc}^2 \,\rho_c \,v_t^3 \,f_2  ,\eqno\autnum $$
where $f_2$ is also a correction factor of order unity to allow for structural
variations;  we use the average density and the radius of the entire convective
region.
\par
Models of Langer (1992, priv.comm.) provide the input together
with data from the textbook of Cox \& Giuli (1968):

$$R\;=\;9.0 \, R_{\odot}\,({M \over {40 \, M_{\odot}}})^{0.512}  ,
\eqno\autnum $$

$$M_{cc}/M\;=\;0.628 \,({M \over {40 \, M_{\odot}}})^{0.466}  ,
\eqno\autnum $$

$$R_{cc}/R\;=\;0.384 \,({M \over {40 \, M_{\odot}}})^{0.377}  ,
\eqno\autnum $$
where $R$ and $M$ are the stellar radius and mass, respectively.  This leads to
a magnetic field of

$$B\;=\;1.9 \,10^6 {\alpha}^{1/4}\,f_1^{1/2} \,f_2^{-1/6} \,({M \over {40 \,
M_{\odot}}})^{-0.058}  \; {\rm Gauss}, \eqno\autnum $$
where $\alpha$ is the fraction of critical rotation at the surface, assumed to
be solid body rotation.  The same argument for Wolf Rayet stars (using data
of Langer 1989, with Helium mass fraction $Y=1.$) gives

$$B\;=\;2.3 \,10^7 {\alpha}^{1/4}\,f_1^{1/2} \, f_2^{-1/6} \,({M_W \over {5 \,
M_{\odot}}})^{0.035}  \; {\rm Gauss}, \eqno\autnum $$
\par
Hence this readily produces magnetic fields, dependent on the rotation rate of
the star, of up to $2 \; 10^6 \; \rm Gauss$ for O stars, and about $10$ times
more for Wolf Rayet stars.  In both cases the induced magnetic field is
nearly independent of stellar mass.  Critical rotation here means that the
convective core has to rotate at an angular velocity which would correspond at
the surface to critical rotation there; but in fact, as we will see below, it
is
not required that the actual surface rotates this fast.
\par
In conclusion we find that we rather easily generate magnetic fields at levels
deep inside the star beyond the local virial limit at the surface of the star,
which gives of the order of a few times $10^4$ Gauss nearly independent of the
rotation rate (Maheswaran and Cassinelli 1988, 1992).  Rotationally induced
circulations can carry these magnetic fields to the surface of the stars on a
time scale much shorter than the main sequence life time.  In this
transport the magnetic field is weakened by flux conservation and so surface
fields in the range $10^3$ to $10^4$ Gauss are quite plausible, and are below
the local virial theorem limit.  Such values are all we require in the
following.  These estimates are valid for massive stars and extend clearly to
below the mass range where we have Wolf Rayet stars as an important final phase
of evolution.  We thus expect many of the arguments in the following to hold
generally for all massive single stars with extended winds, whether slow or
fast, whether red or blue supergiant preceding the supernova explosion. \par
There is a further consequence for the generation of magnetic fields in white
dwarfs:  The most massive stars that do not become supernovae, but white
dwarfs,
are sufficiently massive to contain also convective cores which get exposed
when
the white dwarf is formed.  Thus there ought to be a correlation between
massive
white dwarfs and the detection of strong magnetic fields; this is consistent
with the observational data (Liebert 1992, priv.comm., Schmidt et al. 1992).

\titlea {Magnetically driven winds}

In the standard version of the fast magnetic rotator theory (Hartmann
and MacGregor 1982, Cassinelli 1993) it is assumed that the magnetic field is
radial close to the stellar surface and then starts bending at the critical
point where the radial Alfv\'en velocity, the local corotation velocity and the
wind velocity all coincide.  This produces a long lever arm for the loss of
angular momentum, and is the essence of the criticism of Nerney and Suess
(1987) against the model.  The spindown of fast magnetic rotator Wolf-Rayet
stars is a topic addressed by Poe, Friend and Cassinelli (1989), where it is
again assumed that the field is radial close to the stellar surface.
\par
However, the magnetically driven winds do not require the magnetic field to be
radial near the surface.  In a convection zone near the surface, it is
plausible
to assume that the magnetic field is nearly isotropic in its turbulent
character below the region where the wind gets started, and radial in the
wind zone near to the star, leaving the radial magnetic field lines dominant
for
as long as the flow velocity is below the radial Alfv\'en speed.
\par
However, in a star, where the outer layers are radiative,
it is not clear at all that the magnetic field is initially radial.  Even the
slightest differential rotation will tend to make the magnetic field which
originates in the central convective region to be strongly tangential, and even
the acceleration into a wind does not obviously overturn this tendency
completely.  In fact, the recent model calculations of Wolf Rayet star winds by
Kato and Iben (1992) demonstrate that the critical point of the wind where the
wind becomes supersonic is already inside the photosphere, and so it is
reasonable to suppose that the other critical point where the wind speed
exceeds
the radial Alfv\'en velocity may also be inside the star, if there is such a
point at all - the radial flow velocity may be faster than the radial Alfv\'en
velocity throughout.
\par
In the following we propose to discuss such a wind,
generalizing from Parker (1958) and Weber and Davis (1967).  As in Weber and
Davis we limit ourselves here to the equatorial region.  We use
magneto-hydrodynamics and Maxwells equations and thus have for the angular
momentum transport $L_J$:

$$L_J\;=\;r\,v_{\phi}\,-\,({{B_r} \over {4\,\pi\,\rho\,v_r}})\,r\,B_{\phi}
\;=\;const \;. \eqno\autnum $$
Here the components of the magnetic field are $B_{\phi}$ and $B_r$,
and the components of the wind velocity are $v_{\phi}$ and $v_r$, while
$\rho$ is the density and the index $s$ refers to the stellar surface.
\par
Introducing for the magnetic flux

$$F_B\;=\;r^2\,B_r\,=\;r_s^2\,B_{rs}\;=\;const,\eqno\autnum $$
with flux freezing

$$r \,(v_r B_{\phi} - v_{\phi} B_r) \;=\;- \Omega \,F_B \;=\; const \, ,
\eqno\autnum $$
and mass flux

$$\dot{M}\;=\;4\,\pi\,\rho\,r^2\,v_r\;=\;4\,\pi\,
\rho_s\,r_s^2\,v_{rs}\;=\;const
,
\eqno\autnum $$
the tangential velocity can be written as

$$v_{\phi}\;=\;{{L_J} \over {r}}\,({{1\,-\,{F_B^2 \over \dot{M}}\,{\Omega \over
L\,v_r}} )/( {{1\,-\,{F_B^2 \over \dot{M}}\,{1 \over {r^2\,v_r}}}}})
.\eqno\autnum $$
\par
The angular momentum loss can be written as

$$L_J\;=\;\epsilon \, \Omega\,r_s^2.\eqno\autnum $$
If there is no Alfv\'en critical point outside the star, then
obviously

$$\epsilon \;<\,1.\eqno\autnum $$
\par
We note that the term $F_B^2 / (\dot{M} v_r r^2)$ appearing in the
expression for the tangential velocity can be rewritten with the surface radial
Alfv\'en Mach number

$$M_{Ars}\;=\;v_{rs} /v_{Ars}, \eqno\autnum $$
where

$$v_{Ar}\;=\;B_r / ({4\,\pi\,\rho})^{1/2} ,\eqno\autnum $$
as

$${F_B^2 \over \dot{M}}\,{1 \over {v_r r^2}}\;=\;{1 \over M_{Ars}^2}\,{v_{rs}
\over v_r}\,{r_s^2 \over r^2}\;=\;{1 \over M_{Ar}^2}.\eqno\autnum $$
Similarly we have the relationship

$${F_B^2 \over \dot{M}}\,{\Omega \over L_J\,v_r} \;=\;{1 \over
M_{Ar}^2}\,{r^2 \over r_s^2}\,{1 \over \epsilon}\;=\;{1 \over
M_{Ars}^2}\,{v_{rs}
\over v_r}{1 \over \epsilon}.\eqno\autnum $$
Thus the tangential velocity can be written as

$$v_{\phi}\,=\;{L_J \over r}\,({1\,-\,{v_{rs} \over
{M_{Ars}^2\,
\epsilon\,v_r}}}
)/( {1\,-\,{1 \over M_{Ar}^2}}).\eqno\autnum $$
\par
We can reasonably assume that the radial velocity is steadily increasing
with radius.  The tangential velocity should neither be negative nor exceed the
rotational velocity of the star itself, and so we derive the conditions

$$M_{Ars}\;>\;1 ,\eqno\autnum $$
and

$$M_{Ars}^{-2}\;<\;\epsilon\;<\;1 \eqno\autnum $$
for the conditions envisaged here, that there is no Alfv\'en critical point
outside the star.  These conditions translate into

$${1 \over M_{Ars}^2}\,{v_{rs} \over v_r}{1 \over \epsilon}<\;1 ,\eqno
\autnum
$$
and

$${1 \over M_{Ar}^2}\;<\;1.\eqno\autnum $$
It follows, e.g., that $M_{Ar} \, \sim \, r$ asymptotically.
\par
We define $U_{\epsilon}$ and $U_M$ and obtain

$$0\,<\,U_{\epsilon}\;=\;1\,-\,{{\epsilon r_s^2} \over {r^2}}\,<\,1,
\eqno\autnum
$$ and

$$0\,<\,U_{M}\;=\;1\,-\,{1 \over M_{Ar}^2}\,<\,1.\eqno\autnum $$
The tangential magnetic field can then be written as

$$B_{\phi}\;=\;-{{F_B\,\Omega} \over {r\,v_r}}\,({1\,-\,{{\epsilon\,r_s^2}
\over
{r^2}}})/({1\,-\,{1 \over M_{Ar}^2}}) ,\eqno\autnum $$
and is thus also without change of sign outside the star.  It is easy to verify
that no magnetic flux is transported to infinity (see Parker
1958).
\par
The radial momentum equation is

$$\eqalign{&(v_r {d \over dr} v_r)\,(1-{c_s^2 \over v_r^2}) \;=\;\cr
&2\,{c_s^2 \over r}\,-\,{{G\,M_{\star}} \over r^2}\,+\,{{F_{rad}\sigma_T N}
\over {m_p c}}\,+\,\cr
&{v_{\phi}^2 \over r}\,-\,{1 \over {8\,\pi\,\rho\,r^2}}\,{d \over {dr}}\,(r
B_{\phi})^2.\cr} \eqno\autnum $$
In this equation the radiation force (flux $F_{rad}$) on both lines and
continuum is given with the correction factor $N$ over the Thompson
cross-section, also including the effect due to the chemical element
composition
being different from pure hydrogen; this factor $N$ depends on the physical
state of the gas.  The adiabatic speed of sound is $c_s$.  We here proceed to
evaluate the last term with the expressions already derived and discuss it in
the context of the momentum equation.
\par
The gradient term of $(r \, B_{\phi})^2$ in the momentum equation can then be
rewritten as (on the right hand side of the momentum equation)

$$\eqalign{& (v_r {d \over {dr}} v_r)\,{1 \over M_{Ars}^2} \,({{r_s \Omega}
\over
v_{rs}})^2\,({v_{rs} \over v_r})^3 \, {U_{\epsilon}^2 \over U_M^3} \cr
&-2\,{1 \over M_{Ars}^2}\, r_s \Omega^2 \,({r_s \over r})^3 \,{v_{rs} \over
v_r}\,  {U_{\epsilon} \over U_M^3} \, (\epsilon\,-\,{1 \over
M_{Ars}^2}\,{v_{rs} \over v_r}). \cr} \eqno\autnum $$
A second term on the right hand side of the momentum equation is the
centrifugal force, which can be written as

$$r_s \Omega^2 \,({r_s \over r})^3\,{1 \over {U_M^2}}\,
(\epsilon \,-\,{1 \over {M_{Ars}^2}}\,{v_{rs} \over v_r})^2
.\eqno\autnum $$
The last term in brackets never goes through zero outside the
star because of the conditions we have set above for $\epsilon$ and $M_{Ars}$.
Comparing now the centrifugal term with the second term from the gradient of
the
tangential magnetic field, we note that asymptotically these two terms differ
by
the factor

$$-{1 \over 2}\,(\epsilon M_{Ars}^2)\,{v_r \over v_{rs}},\eqno\autnum $$
where $\epsilon M_{Ars}^2 \,>\,1$ and $v_r /v_{rs}$ likely to exceed a value of
$2$ at large radii $r$; hence the centrifugal term, which provides some
acceleration, is likely to dominate over the other term at large $r$.  Here we
have ignored all the terms of order unity that approach unity with both the
radius $r$ and the radial velocity $v_r$ becoming large.  Even close to the
star
the centrifugal force may dominate.
\par The first term in the gradient of
the
tangential magnetic field is more interesting, however.  This term becomes a
summand to the various terms multiplying $(v_r {d \over {dr}} v_r)$ and has
there, on the left hand side, a minus sign and is to be compared with unity in
the case that we are already at supersonic speeds, as argued by Kato and Iben
(1992).
\par
We define an Alfv\'en Machnumber with respect to the total
magnetic field with

$$M_A^2 \;=\;{{v_r^2 4 \pi \rho} \over {B_{\phi}^2 + B_r^2}}. \eqno\autnum $$
Generally we have obviously

$$M_A \;<\;M_{Ar} \;. \eqno\autnum $$
With this generalized Alfv\'en Machnumber we can
rewrite the entire factor to $(v_r {d \over {dr}} v_r)$ in a simple form

$$1\,-\,{1 \over M_s^2}\,-\,({M_{Ar}^2 \over M_A^2} - 1)/(M_{Ar}^2 - 1) ,
\eqno\autnum $$
where $M_s$ is the sonic Mach number for the radial flow velocity.  Critical
points appear whenever this expression goes through zero or a singularity.
This expression is easily seen to be equivalent to eq.(6) of Hartmann and
MacGregor (1982); they, however, assumed that the magnetic field is initially
radial and neglected radiative forces.  This shows
that we have the critical point of Weber and Davis (1967) again for
${1 \over M_s^2}\, \ll \, 1$,
when and if $M_{Ar}=1$, but we have in our case another critical point at
$M_A=1$, which is the fast magnetosonic point (see Hartmann and MacGregor
1982).  This critical point may never appear for certain choices of the
parameters.  In order to make a realistic judgement on this question, we would
have to combine the model of Kato and Iben (1992) with a realistic treatment of
the magnetic field inside the star, so that we can derive the range of possible
properties of the magnetic field near the surface of the star.
\par
However, there are a few general conclusions one can draw already:   If the
magnetic field inside the star begins as a dominantly tangential field, then
it is by no means clear whether there is any point at which we have $M_{Ar}<1$,
either inside or outside the star; going outwards the unwinding of the magnetic
field is intimately coupled to the initially weak radial flow, and so
$M_{Ar}>1$
may hold throughout.  In that case the term derived from the radial gradient of
the tangential magnetic field goes through unity only when the generalized
Alfv\'en Mach number goes through unity. If, as we argued, $M_{Ar}>1$
possible throughout, then, again going outwards with radius, we start with a
low
generalized Alfv\'en Mach numbers $M_A$, either supersonic or subsonic flow,
the
entire expression is negative, matching the negative gravitational
force on the right hand side.  Far out, both sonic and generalized
Alfv\'en Machnumbers are large, and so the expression is positive.  It follows
that the expression has to go through zero.  The condition $M_A=1$ here is the
fast magnetosonic point, and the radial velocity there corresponds to the
Michel-velocity.  Or, in other words, the radial velocity is equal to the total
Alfv\'en velocity.
\par
We know from observations that the winds in OB and Wolf Rayet stars are
very strongly supersonic; using our model we deduce that the radial velocity is
weakly super-Alfv\'enic with respect to the tangential (dominant) magnetic
field
component, and so we have far outside the star a configuration where $M_s>>1$,
$M_{Ar}>>1$ and $M_A
\simge 1$, but this latter condition may not necessarily
be
satisfied by much.  This means that the expression is near to unity and
somewhat
smaller than unity.  This entails the condition that the right hand side has to
be positive, which is readily interpreted as possibly arising from line
radiation, because that term is the only one which has the same radial
dependence as the gravitational force.  Comparing then the effect of line
driving (Lucy and Solomon 1970, Castor et al. 1975) with and without such a
magnetic field, the net effect is an amplification in the sense that for
$M_{Ar}^2 \gg 1$ the velocity gradient is asymptotically increased by

$$1/(1\,-\,{1 \over M_A^2}).\eqno\autnum $$
We thus argue that there is a magnetic field configuration where the field
is mostly tangential already inside the star, and remains mostly tangential
outside the star, where the initial acceleration of the wind is done by the
opacity mechanism discussed by Kato and Iben (1992); outside the star we have
a line-driven wind, but the amplification by the effect of the tangential
magnetic field really produces the large momentum in the wind.
\par
To see the properties of the equations better, we simplify by dropping all
right hand terms in the differential equation except for the gravitational and
the radiative force, and introduce characteristic length and velocity scales

$$r_{\star} \;=\;{{L N} \over L_{edd}}\,({{G M} \over v_{rs}^2}) \,
({v_{rs} \over v_{Ars}})^{4/3} \, ({v_{rs} \over {r_s \Omega}})^{4/3} ,
\eqno\autnum $$
and

$$v_{\star} \;=\; v_{rs} \,({v_{Ars} \over v_{rs}})^{2/3} \, ({{r_s \Omega}
\over v_{rs}})^{1/3} ,\eqno\autnum $$
which is easily recognized as the Michel velocity (Hartmann and MacGregor
1982).  The Eddington luminosity is given by

$$L_{edd} \;=\;{{4 \pi G M m_p c} \over \sigma_T}.\eqno\autnum $$
In the approximation that $U_{\epsilon} \, = \, U_M \, \simeq\,1$ and $r
\Omega \gg v_r$ the Michel velocity corresponds to the overall Alfv\'en
velocity.
The differential wind equation then reads in dimensionless length  $x$ and
velocity $y$

$$(1 - {1 \over y^3}) \,{{d y^2} \over {d \, (1/x)}} \;=\; -f_{rad} ,
\eqno\autnum $$
with

$$f_{rad}=(1 - {L_{edd} \over {N L}}).\eqno\autnum $$
Clearly we require then that the radiative force dominates, by however little,
to obtain the correct sign of the right hand side, thus $f_{rad} \;> \,0$.  The
analytic solution yields then two extreme possibilities, either

$$v_{r \infty} \;=\; v_e \, ({{N L f_{rad}} \over L_{edd}})^{1/2} ,
\eqno\autnum $$
or

$$v_{r \infty} \;=\; \sqrt{2} \,({v_{Ars} \over v_{rs}})^{1/3} \, ({{r_s \Omega}
\over v_{rs}})^{1/3} \,v_{\star} , \eqno\autnum $$
where we have used the definition

$$v_e \;=\;({{2 G M} \over r_s})^{1/2} \eqno\autnum $$
for the surface escape velocity from the star. Obviously, we require that

$$2^{3/2} \,({v_{Ars} \over v_{rs}}) \, ({{r_s \Omega}
\over v_{rs}}) \; > \; 1 \, ,\eqno\autnum $$
in order to have a final velocity above the Michel velocity.  We note that

$${{\Omega r_s} \over v_{rs}} \;=\;{B_{\phi s} \over B_{r s}} \, {U_{M s} \over
U_{\epsilon s}} , \eqno\autnum $$
and thus the condition is indeed well fulfilled for a magnetic field
configuration which is highly tangential at the stellar surface.  By the same
condition, the initial surface velocity is below the Michel velocity (except
for
the factor $2^{3/2}$ which is of order unity).  This is necessary for the flow
to have a transition through the fast magnetosonic point.  This sharpens the
approximations introduced above.
\par
We have thus in this approximation two possible extreme solutions:  First, for
small magnetic field, the Alfv\'en velocity drops out and the line driving is
the regulating agency; second, for large magnetic field, the terminal wind
velocity is not far from the tangential Alfv\'en speed, and the line driving
effect is small, except in that it provides the source of momentum to be
amplified.  This latter picture is the concept we propose to explain the
momentum in the winds of Wolf Rayet stars.  This is different from the
solutions
of Hartmann and MacGregor (1982) in the sense, that in their wind solutions the
main acceleration of the wind is between the slow magnetosonic point and the
Alfv\'en point, both of which are not outside the star in the case which we
discuss; our solutions are similar, however, in the sense that both in their
case as here, the terminal velocity of the wind in the case of interest is not
far from the the total Alfv\'en velocity.  In this picture the angular momentum
of the star lost refers to a characteristic level inside the star and so the
criticism of Nerney and Suess (1987) is countered, the angular momentum loss of
the star through mass loss is minimized.
\par
Using the observed wind velocities then gives an estimate for the Alfv\'en
velocity, and thus implies - assuming our wind driving theory to be correct -
that the magnetic field is quite high, and of the order of $3$ Gauss at the
fiducial radius of $10^{14}$ cm.  Or, to turn the argument around, the magnetic
field strengths implied by our cosmic ray arguments provide a possible
explanation for the origin of the momentum of Wolf Rayet star winds.  On the
surface of the star the magnetic field strength implied is of order a few
thousand Gauss, quite easily within the limits implied by the surface virial
theorem.  Also the surface of the star may not rotate as fast, but the inside
could still rotate at an angular velocity corresponding to near critical at the
surface.  We emphasize that here a proposed wind driving theory provides an
independent argument for the magnetic field strength on the surface of Wolf
Rayet stars.

\titlea {The nonthermal radio emission}

In this section we derive the basic expressions for the luminosity,
spectrum, and time dependence of the nonthermal radio emission from
single shocks in the winds of massive stars.  In the subsequent sections
we then use these expressions to discuss radiosupernovae, young radio
supernova remnants in starburst galaxies, Wolf Rayet stars, and OB
stars.
\par
Electrons suffer massive losses due to Synchrotron radiation and so they
can achieve only energies up to that point where the acceleration and loss time
become equal; this is a strong function of latitude since the magnetic field
varies rather strongly with latitude.  The shock transfers a fraction $\eta$ of
the bulk flow energy into relativistic electrons.  The emissivity of an
electron
population with the spectrum

$$N(\gamma) \; d\gamma \;=\; C \;\gamma^{-p} \;d\gamma \eqno\autnum $$
is given by (Rybicki and Lightman 1979) in cgs units for $p=2.47$

$$\epsilon_{\nu}\;=\; 1.05\;10^{-17}\; C\;B^{1.735}\;\nu^{-0.735} \;{\rm erg
\,sec^{-1} \,cm^{-3} \,Hz^{-1}}.  \eqno\autnum $$
and for $p=7/3$ by

$$\epsilon_{\nu}\;=\; 3.90\;10^{-18}\; C\;B^{5/3}\;\nu^{-2/3} \;{\rm erg
\,sec^{-1} \,cm^{-3} \,Hz^{-1}}.  \eqno\autnum $$
Similarly the absorption coefficient for synchrotron self absorption is
($p=2.47$)

$$\kappa_{\nu,syn} \;=\;3.56 \,10^{12}\,C\,B^{2.235}\, \nu^{-3.235} \,\rm
cm^{-1} ,\eqno\autnum $$
and for $p=7/3$

$$\kappa_{\nu,syn} \;=\;1.23 \,10^{12}\,C\,B^{13/6}\, \nu^{-19/6} \,\rm
cm^{-1}.\eqno\autnum $$
In all these expressions we have averaged the aspect angle.  We use here
at first the case of a wind speed equal to the shock velocity and the limit of
a
strong shock in a gas of adiabatic index $5/3$, which leads to an optically
thin
synchrotron spectrum of $-0.735$.  We will use the limit of large shock speeds
below for the discussion of the radiosupernovae, where the radio spectral index
is $-2/3$.  The free-free opacity is given by

$$\kappa_{\nu,ff} \;=\;0.21 \,T_e^{-1.35} \,\nu^{-2.1} \,n_e^2 \,\rm cm^{-1}
, \eqno\autnum $$
in an approximation originally derived by Altenhoff et al. (1960) and widely
disseminated by Mezger and Henderson (1967).  Here $T_e$ and $n_e$ are the
electron temperature and density, and the approximation has been used that
the ion and electron density are equal, and that the effective charge of the
ions is $Z\,=\,1$.  We note that for cosmic abundances we have the following
approximations (Schmutzler 1987) assuming full ionization:

$$\eqalign{& \rho \;=\;1.3621 \,m_H \;n , \cr
& n_e \;=\; 1.181 \,n , \cr
& n_i \;=\; 1.086 \,n , \cr} \eqno\autnum $$
where $n$ is the total Hydrogen density (particles of mass $m_H$ per cc), and
$n_i$ the ion density.  With these approximations for full ionization the
free-free opacity would increase a factor of two over the approximation by
Altenhoff et al. (1960), which, however, is compensated by the incomplete
ionization (see, again, the calculations by Schmutzler 1987) in the photon
ionized regions (here, winds) near massive stars.  In the following we will use
the approximation by Altenhoff et al. replacing $n_e\,=\,n$.
\par
The synchrotron luminosity in the optically thin limit is then given by an
integration over the emitting volume, which we take in the context of our
simplified picture to be a spherical shell of thickness $r/4$, for
$U_1/V_W\,=\,1$ decreased to $r/8$.  The radio emission only arises from that
part of the shell where the shock velocity is larger than the local Alfv\'en
velocity and where the  synchrotron loss time is longer than the local
acceleration time.  These conditions can lead to a restricted range in latitude
for the emission.  The luminosity is then given by the integral over the
latitude dependent emissivity.  Here we have to normalize the cosmic ray
electron density to the shock energy density, using our induced latitude
dependence.  In paper CR I we demonstrated that depending on the orientation of
the magnetic field the $\theta$-dependence is either peaked toward the poles,
when the drift is toward the poles, or peaking near the equator, when the drift
is toward the equator.  Clearly, the second case produces very much stronger
radio emission, since the magnetic field is also strongest near the equator,
and
so we will concentrate on this case and discuss the opposite case briefly at
the
end.  Using a lower bound for the electron spectrum much below the rest mass
energy and an upper bound much above that value we have for the constant $C$
then the expression

$$C(\mu)\;=\; C_o \;(1-\mu^2)^{3/2} \eqno\autnum $$
with

$$\mu_{\star} \, > \,\mu \, > \, 0 \eqno\autnum $$
where $\mu_{\star}$ refers to that latitude where the latitude dependent
acceleration breaks down:

$$(1-\mu_{\star}^2)^{1/2}\;=\;{3 \over 4}\; {U_1 \over c} \; \ll \; 1 \; ,
 \eqno\autnum $$
from our argument about the knee energy (see section 8 of paper
CR I) for protons and other nuclei.  At the equator the energy density of the
electrons can be written as

$$C_o\;=\;(p-2) \eta \; \rho U_1^2 /(m_e c^2) \eqno\autnum $$
where we use just the relativistic part of the distribution function and
incorporate the uncertainty on the existence and strength of any
subrelativistic part of the electron distribution function in the factor
$\eta$.  The integration over latitudes leads to an integral correction
factor of

$${1 \over 2} \,B({{p+11} \over 4},{1 \over 2}) \;\;,$$
where $B(z_1,z_2)$ is the Beta-function, and $p$ the powerlaw index of the
electron distribution function.  For the two powerlaw indices used here,
$p=7/3$
and $p=2.470$, this integral is both very close to $0.50$ in value.  We also
have

$$\rho\;=\;{{\dot{M}} \over {4 \pi r^2 V_W}}. \eqno\autnum $$
Assuming that all latitudes contribute up to $\mu_{\star}$ assumed to be close
to unity, the final expression for the luminosity is then

$$\eqalign{L_{\nu}(nth)\;=&\;4.0\,10^{24}\;\rm erg/sec/Hz \cr
& \;\eta_{-1} \, ({\dot{M}_{-5} \over V_{W,-2}})\; U_{1,-2}^2 \;
B_{0.5}^{1.735}\;r_{14}^{-0.735}\;\nu_{9.7}^{-0.735} \cr}\;\;. \eqno\autnum $$
Here the mass loss is in units of $10^{-5}\; \rm M_{
\odot} /yr$, the
unperturbed
magnetic field strength at the reference radius of $10^{14} \, \rm cm$ in units
of of $3 \,\rm Gauss$, the wind and the shock velocity in units of
$0.01 \;c$, and as reference frequency we use $5$ GHz.  For the powerlaw index
of
$p=7/3$ this luminosity is

$$\eqalign{L_{\nu}(nth)\;=&\;8.1\,10^{24}\;\rm erg/sec/Hz \cr
& \;\eta_{-1} \,{\dot{M}_{-5} \over V_{W,-2}}\; U_{1,-2}^2 \;
B_{0.5}^{5/3}\;r_{14}^{-2/3}\;\nu_{9.7}^{-2/3} \cr}\;\;. \eqno\autnum $$
\par
Here we note that this emission might become optically thick both to free-free
absorption in the lower temperature region outside the shock, or to to
Synchrotron self absorption inside the shocked region.  In the approximation,
that we use the equatorial region in the slab model (i.e., direct central
axis radial integration), we obtain for the critical radius $r_{1,ff}$, where
the
free-free absorption has optical thickness unity:

$$r_{1,ff}\;=\;1.35 \,10^{14} \, ({\dot{M}_{-5} \over V_{W,-2}})^{2/3} \,
\nu_{9.7}^{-0.70} \rm \,cm. \eqno\autnum $$
We have used here the analytical approximation for the free-free opacity
introduced above and an assumed temperature of $T_e\;=\; 2 \,10^4 \;\rm K$.
\par
Similarly, for synchrotron self absorption the critical radius $r_{1,syn}$ is
given for $p=2.470$ by:

$$\eqalign{r_{1,syn}\;=&\;3.18 \,10^{14} {\, \rm cm} \cr &
\;\eta_{-1}^{0.309}\,({\dot{M}_{-5} \over V_{W,-2}})^{0.309}\,
U_{1,-2}^{0.618} \, B_{0.5}^{0.691} \,\nu_{9.7}^{-1} \cr} \;\;. \eqno\autnum $$
and for $p=7/3$:

$$\eqalign{r_{1,syn}\;=&\;3.99 \,10^{14} \, {\rm cm} \cr &
\;\eta_{-1}^{0.316}\,({\dot{M}_{-5} \over V_{W,-2}})^{0.316}\, U_{1,-2}^{0.632}
\, B_{0.5}^{0.684} \, \nu_{9.7}^{-1} \cr} \;\; . \eqno\autnum $$
The maximum synchrotron luminosity is given by the dominant absorption process;
which one is stronger is given by comparing the relevant radii; in the
numerical
example synchrotron self absorption happens to be stronger.  Free-free
absorption is dominant for $p=2.470$ if

$${\dot{M}_{-5} \over V_{W,-2}} \;>\;11.0 \,\eta_{-1}^{0.864}\,U_{1,-2}^{1.728}
\,  B_{0.5}^{1.932} \,\nu_{9.7}^{-0.839}, \eqno\autnum $$
and for $p=7/3$ if

$${\dot{M}_{-5} \over V_{W,-2}} \;>\;22.0 \,\eta_{-1}^{0.901}\,U_{1,-2}^{1.802}
\,  B_{0.5}^{1.951} \,\nu_{9.7}^{-0.856}. \eqno\autnum $$
We note that the parameter $\eta$ might well be quite low.
\par
In the following we give then the maximum luminosities calculated by using
the proper optical depth for a central axis approximation
in a slab geometry for the radiative transfer at the equator; this gives an
additional factor of about $2/3$  (see below).  However, for the total
emission we do integrate properly over all latitudes, while for the absorption
we approximate by using the equatorial values.  This is a fair approximation,
since both emission and absorption decrease towards the poles, but the
absorption decreases even faster.  The exact numerical value in our
approximation
is used here.
\par
In the case that free-free absorption dominates, the
maximum luminosity is then given by $p=2.470$

$$\eqalign{L_{\nu}(nth)\;=&\;1.8 \,10^{24} \, {\rm erg\,sec^{-1}\,Hz^{-1}}
\cr
&
\eta_{-1}\,({\dot{M}_{-5} \over V_{W,-2}})^{0.510}\, U_{1,-2}^2 \;
B_{0.5}^{1.735}\,\nu_{9.7}^{-0.221} \cr} \;\;, \eqno\autnum $$
and for $p=7/3$ by

$$\eqalign{L_{\nu}(nth)\;=&\;3.8 \,10^{24} \, {\rm erg\,sec^{-1}\,Hz^{-1}}
\cr
&
\eta_{-1}\,({\dot{M}_{-5} \over V_{W,-2}})^{0.556}\, U_{1,-2}^2 \;
B_{0.5}^{1.667}\,\nu_{9.7}^{-0.200} \cr}\;\;. \eqno\autnum $$
\par
In the case that Synchrotron self absorption dominates, these
maximum luminosities are given by $p=2.470$

$$\eqalign{L_{\nu}(nth)\;=&1.2 \,10^{24} \, {\rm   erg\,sec^{-1}\,Hz^{-1}}
\cr
&
\eta_{-1}^{0.773}\,({\dot{M}_{-5} \over V_{W,-2}})^{0.773}\, U_{1,-2}^{1.546}
\, B_{0.5}^{1.227}
\cr}\;\;, \eqno\autnum $$
and for $p=7/3$ by

$$\eqalign{L_{\nu}(nth)\;=&2.2 \,10^{24} \, {\rm   erg\,sec^{-1}\,Hz^{-1}}
\cr
&
\eta_{-1}^{0.789}\,({\dot{M}_{-5} \over V_{W,-2}})^{0.789}\, U_{1,-2}^{1.579}
\, B_{0.5}^{1.211} \cr} \;\;.\eqno\autnum $$
Here we do not consider mixed cases.
\par
Obviously, the ratio of mass loss rate and wind velocity can be different by
many orders of magnitude among predecessor supernova stars, and their
numerical values will have to be argued on the basis of observations.
\par
It is useful to also calculate the
thermal radio emission, using the same standard parameters, then adjust our
parameters to a realistic range, and then compare the nonthermal luminosities.
The thermal emission can be properly integrated, allowing for the sphericity of
the wind structure (Biermann et al. 1990) to give:

$$L_{\nu}(th)\;=\; 4{\pi}^2\;B_{\nu}(T)\;r_{1,ff}^2\;\Gamma({1 \over 3})
\eqno\autnum $$
With our standard parameters this luminosity is at the frequency of maximum
emission (i.e. what we observe in a steady wind)

$$L_{\nu}(th)\;=\;3.0\;10^{17}\;({\dot{M}_{-5} \over V_{W,-2}})^{4/3} \,
\nu_{9.7}^{+0.60} \;\rm erg/sec/Hz \; ,\eqno\autnum $$
weakly dependent on electron temperature.
\par
In the following we make a number of consistency checks on the basic notions
used above:
\par
First, we note that shocks can only exist when the shock velocity is larger
than the Alfv\'en velocity.  Since in our wind driving theory developed above
the
wind velocity is just a little larger than the Alfv\'en velocity itself, this
implies that the shock velocity in the frame of the flow has to be at least the
same velocity as the wind, and so - within our analytical approximations - we
have the condition that always

$$U_1 \geq V_W , \eqno\autnum $$
where the limit of the equality corresponds to a spectral index for the
Synchrotron emission of $-0.735$ and in the limit of large shock velocity to
$-2/3$.  Here we have to note that the Alfv\'en velocity is colatitude
$\theta$ dependent and is proportional to $sin \,\theta$.  Thus,
even for lower shock speeds, there is a latitude range where a shock can be
formed, but then the luminosity is very much reduced.
\par
Second, we have to check that the Synchrotron loss time is larger than the
acceleration time, because otherwise we would not have any electrons at the
appropiate relativistic energies.  This condition can be rewritten as

$$\nu_{9.7} \, B_{0.5}^3 \; \simle \;  \, U_{1,-2}^2 \,r_{14}.
\eqno\autnum $$
This is the condition at the equator, and the condition gets weaker at
higher latitudes.  This makes it obvious, that for all reasonable ranges of the
parameters acceleration can succeed to the required electron energies.  It also
shows that at higher frequencies or smaller radii the
condition would fail.  At smaller radii the emission is usually optically
thick (see above), and higher frequencies are as yet difficult to observe at
the required sensitivity.  The implied high frequency cutoff in the
observed synchrotron spectra would, however, be an important clue.
\par
Third, we have to check whether the implication that the shock speeds are
typically similar to the wind speeds, is supported by data on Wolf Rayet stars,
and their theoretical understanding.  The wind calculations with shocks suggest
that the typical shock velocities are indeed of order the wind speed itself or
somewhat higher (Owocki et al. 1988), and so we expect a range in radio
spectral
indices of $-0.667$ to $-0.735$ or steeper if the shocks are not strong, i.e.
if $U_1/U_2 \,<\,4$.  We note that the actual value of the nonthermal
luminosity
is changed only moderately in this range of spectral indices.
\par
Fourth, we have to discuss optical thickness effects in more detail:  A
shock travels from the region inside of where free-free absorption dominates
through this region to the outside.  The emission then is first weak and has a
steep spectrum due to the strong frequency dependence of the free-free
absorption and the exponential cutoff induced, then approaches a peak in
emission with a spectral index approaching $-0.67$ to near about $-0.735$ or
steeper as discussed above, and then becomes weaker with the optically thin
spectrum.  At the location where our ray encounters the shock, the temperature
of the gas increases drastically, and so the differential optical depth for
free-free absorption goes to zero.  Hence we have the simple case that we have
emission inside the shock and absorption outside.  We limit ourselves
to the central axis as a first approximation, and also neglect the spatial
variation of the emission itself (see above).  This is equivalent to pure
screen-like absorption, however with a screen which extends from the shock to
the outside and so is variable.  The radio luminosity is given by (using
free-free absorption)

$$L_{\nu}(nth)\;=\;e^{-{\tau_{\nu}}}\;L_{\nu}(nth, \, no \, abs).
\eqno\autnum $$
We have the spectral index $-\alpha_{thin}$ in the optically thin regime, and
the time dependence of the spectral index given by

$$\alpha (\nu)\;=\;-{\alpha_{thin}}\;(1-(t^{\star}_{\nu}/t)^3). \eqno\autnum $$
The spectral index is positive and very steep at first and then goes through
zero to become slowly negative approaching asymptotically the optically thin
spectral index.  The nonthermal luminosity considered as a function of time
sharply rises at first, then peaks at optical depth $
\alpha_{thin}/3$,
obviously
strongly dependent on frequency, and finally drops off as
$1/t^{\alpha_{thin}}$:

$${{d \,ln \,L_{\nu}(nth)} \over {d \,ln \,t}} \;=\;
-{\alpha_{thin}}\;(1-{10 \over 7}\,(t^{\star}_{\nu}/t)^3). \eqno\autnum $$
Here we see that the time dependence of the flux density at a given frequency
and the time dependence of the spectral index are closely related.  The
time of luminosity maximum depends on frequency as

$$t_{max \, L} \; \sim \; \nu^{-0.7} ,  \eqno\autnum $$
from the frequency dependence of the optical depth

$$\tau_{\nu} \; \sim \;t^{-3} \, \nu^{-2.1}.\eqno\autnum $$
These relationships can be used to check on the importance of free-free
absorption in our approximations.
\par
Synchrotron self absorption is due to internal absorption inside the shell, and
so again in the central axis approximation we have

$$L_{\nu}(nth)\;=\;{{1-e^{-\tau_{\nu}}} \over
{\tau_{\nu}}} \, L_{\nu}(nth, \, no \, abs). \eqno\autnum$$
This then results in the time dependence for the spectral index of

$$\alpha (\nu)\;=\;{5 \over 2}\,+\,{{5+2\alpha_{thin}} \over
2}\,\tau_{
\nu}\,-\,
{{5+2\alpha_{thin}} \over 2}\,{\tau_{\nu} \over {1-e^{-\tau_{\nu}}}} ,
\eqno\autnum $$
with

$$\tau_{\nu}\;\sim\;t^{-(5+2\alpha)/2} \,\nu^{-(5+2\alpha)/2}. \eqno\autnum $$
The time dependence of the double logarithmic derivative of luminosity on time
is exactly the same

$${{d \, ln \, L_{\nu}(nth)} \over {d\, ln \,  t}} \;=\;\alpha (\nu)
. \eqno\autnum $$
It follows that the time of maximum depends on frequency as

$$t_{max \, L} \; \sim \; \nu^{-1}.  \eqno\autnum $$
This is a characteristic feature for synchrotron self absorption in the
approximation used (and well known from radioquasars) and differs from the case
considered above, for free-free absorption.
\par
Thus, the true maximum of the luminosity is given by an optical depth less than
unity, which gives a correction factor to the luminosities introduced above
(eqs. 88 through 91) of about $2/3$; we have corrected the luminosities
introduced there for this factor.
\par
Finally, we have to comment on the sign of the magnetic field.  We have
used here throughout the assumption that the magnetic field is oriented such
that
the drifts are towards the equator for the particles considered.  Clearly,
since we consider stars with a magnetic field driven by turbulent convection in
a rotating system, we can expect that there are sign reversals of the magnetic
field just as on the Sun.  For the other sign of the magnetic field and the
other drift direction there is by many powers of ten less nonthermal emission
and so it is to be expected that at any given time we should detect at most
half
of the stars in nonthermal radio emission; occasionally we might even catch a
shock travelling through the region in the wind where the sign is reversing
itself, because the wind mirrors the time history of the star in terms of
magnetic field.
\par
In the following we will first discuss the radio observations of supernovae,
then Wolf Rayet stars, and finally OB stars.

\titlea {Radiosupernovae}

The adopted value for the magnetic field, however, was derived from the notion
that the subsequent supernova shocks accelerate particles to extremely high
energies.  This argument can be checked with the observations of those
supernovae of which the radioemission in the wind was really observed, five
sources discussed by Weiler and colleagues in a number of papers (Weiler et al.
1986, 1989, 1990, 1991, Panagia et al. 1986) and the supernova 1987A (Turtle et
al. 1987, Jauncey et al. 1988, Staveley-Smith et al. 1992) and related radio
sources in starburst galaxies. However, we restrict ourselves to those
supernovae for which we can reasonably assume that the predecessor star was
indeed a Wolf Rayet star, and this we will do using statistical arguments  in
the subsequent section.
\par
In fact, our model can be paraphrased as a numerical version of Chevaliers
(1982)
model with the parameters fixed:  In the screen approximation valid for
free-free absorption (see above for details), the time evolution of the
nonthermal emission in Chevaliers model can be written as

$$L_{\nu}(nth)\;=\;K_1\,\nu^{\alpha} \,t^{\beta}\,e^{-K_2\,t^{\delta}}
 \eqno\autnum $$
with $\alpha$, $\beta$ and $\delta$ to be fitted to the data.  For fast strong
shocks our model predicts that $\alpha \,=\,\beta\,=\,-2/3$ and
$\delta\,=\,-3$, and for slower shocks that still $\alpha\,=\,\beta$ but
larger in number, for $U_1/V_W\,=\,1$, $\alpha\,=\,\beta\,=\,-0.735$, for
instance.  This is very close to the detailed fits for the four supernovae
listed above (Weiler et al. 1986).
\par
For SN 1986J the data clearly cover a sufficiently large time to test this
numerical model in more detail; Weiler et al. (1989) find that this simple
model
in its screen approximation does not provide a good fit to the early epochs.
We
suspect similar to Weiler et al., that this lack of a good fit can be traced to
the simplification that we consider the external absorption as a simple screen,
and disregard the lateral structure.  Further possible reasons for a failure to
strictly adhere to the simplified model are the following:  a) The pre-shock
wind may not be smooth in its radial behaviour, there might have been a weaker
shock running through earlier which nevertheless can disturb the radial density
profile (see the calculations by MacFarlane and Cassinelli 1989).  b) Mixing
between free-free absorption (outside the shock region) and synchrotron
self-absorption (inside the shock region).  c)  The structure of acceleration
in
its latitude dependence is considered here only for the acceleration of
particles (see paper CR I), but not for absorption and radiative transfer.
Given a very detailed multifrequency data set it would be interesting to model
the data fully.
\par
In fact, we can use the numerical values for the radii for maximum
luminosity derived above to obtain the pre-shock wind density, which is
proportional to $
\dot{M}/V_W$.  With the data given by Weiler et al. (1986)
this
yields for the supernova 1979c

$${\dot{M}_{-5} \over V_{W,-2}}(1979{\rm c}) \;=\; 3.0 \,10^3 \,
U_{1,-1.5}^{3/2},$$
and for the supernova 1980k

$${\dot{M}_{-5} \over V_{W,-2}}(1980{\rm k}) \;=\; 3.4 \, 10^2 \,
U_{1,-1.5}^{3/2}.$$
Clearly, these numbers tell us that the predecessor stars had a slow wind, and
thus were probably red supergiants (see Weiler et al. 1986). This then implies
for shock speeds of $0.03 \,c$, $\eta=0.1$, and free-free absorption being
dominant, that the nonthermal luminosities at $5$ GHz expected versus observed
are

$$L_{max}(1979{\rm c})\;=\;3.0 \,10^{27} \;{\rm versus}\; 2.\,10^{27} {\rm erg
\,sec^{-1}\,Hz^{-1}} ,$$
and

$$L_{max}(1980{\rm K})\;=\;9.0 \,10^{26} \;{\rm versus}\; 1.\,10^{26} {\rm erg
\,sec^{-1}\,Hz^{-1}} ,$$
both for the assumed strength of the magnetic field.  Since the luminosity is
proportional to the magnetic field strength to the power $5/3$, and directly
proportional to the electron efficiency parameter $\eta$, we derive thus a
lower limit to the magnetic field, given an upper limit on $\eta$.  Since
$\eta=0.1$ is unlikely to be surpassed by much, using the ratio of the
luminosities expected/observed of the two cases above yields an estimated
lower limit to the magnetic field strength at our reference radius of

$$B \; > \;1.5 \,\eta_{-1}^{-3/5} \, U_{1,-1.5}^{-6/5} \,{\rm Gauss}.$$
This implies a strong lower limit to the magnetic field from using $\eta=1$ of
$0.4 \, \rm Gauss$.  The uncertainty in these estimates is clearly at least a
factor of $2$.  For 1987A, the initial radio luminosity was indeed of order
$10^{25}$ erg/sec/Hz and so, applying our model with a shock speed of order
$0.03$ c is consistent with our expectation of then $3.4 \,10^{25} \, \eta_{-1}
$
erg/sec/Hz; we do not wish to discuss 1987A in any detail here.
\par
This demonstrates as argued in section 7, that strong magnetic fields also
exist in the winds of massive stars in the red part of the Hertzsprung-Russell
diagram, where the winds are slow; note that the predecessor to supernova 1987A
was a blue supergiant with a fast wind.
\par
The only unknown parameter in all these predictions is the efficiency of
electron acceleration $\eta$ and the strength of the magnetic field; with
$
\eta$
close to $0.1$, clearly close to the maximum number reasonable, leads then to
the
requirement that the magnetic field strength is near to what we assumed, $3$
Gauss at $10^{14}$ cm, but even higher, if $\eta$ is very much less than unity.
This again confirms independently that indeed the magnetic field has to be as
high as argued by Cassinelli (1982, 1991) to drive the winds and as is required
to accelerate cosmic rays particles to energies near $3 \,10^9$ GeV.
\par
For supernova predecessor stars with fast winds like Wolf Rayet and OB
stars, it is of interest to ask whether the expected luminosity violates the
Compton limit, famous from the study of radioquasars.  At the Compton limit the
first order inverse Compton X-ray luminosity becomes equal to the synchrotron
luminosity, and it has been found from observations that compact radioquasars
are close to this limit and indeed have strong X-ray emission.  This question
can be formulated as a limit to the brightness temperature of the radio source
(using here  synchrotron self absorption) which then gives the limit

$$\eta_{-1} \,({\dot{M}_{-5} \over V_{W,-2}}) \,U_{1,-1.5}^2 \,B_{0.5}^{-1} \,
\nu_{9.7}^{1.27} \;<\;1.5 \,10^7.\eqno\autnum $$
For free-free absorption the limit is

$$\eta_{-1} \,({\dot{M}_{-5} \over V_{W,-2}})^{-0.777} \,U_{1,-1.5}^2
\,B_{0.5}^{5/3} \, \nu_{9.7}^{-0.6} \;<\;0.18.\eqno\autnum $$
The first of these conditions is almost certainly always fulfilled, while the
second one may be so tight as to suggest that inverse Compton X-rays might be
observable.  This has indeed been checked with modelling successfully the
observed X-ray spectra beyond photon energies of $2$ keV by Chen and White
(1991a) for Orion OB stars.  This suggests that the inverse Compton X-ray
luminosity ought to be less, usually considerably less than the Synchrotron
luminosity.  Most of the X-ray emission from hot stars is thought to arise from
those same shocks in free-free X-ray emission which we consider for particle
acceleration; a modelling of this was done by White and Long (1986) and
MacFarlane and Cassinelli (1989).

\titlea {The statistics of Wolf Rayet stars and supernovae}

There are a variety of ways to estimate the relative frequency of Wolf Rayet
star supernova explosions relative to supernova explosions of lower mass
stars (Hidayat 1991, Leitherer 1991, Massey and Armandroff 1991, Shara et al.
1991).
\par
In our Galaxy there are between 300 and 1000 Wolf Rayet stars, which have
an average lifetime of about $10^5$ years.  This gives an estimated occurrence
of Wolf Rayet supernovae of about one every 100 to 300 years.  Since the total
rate of supernovae in our Galaxy is estimated at about one every 30 years, this
means that roughly one in 3 to one in 10 supernovae ought to represent the
explosion of a Wolf Rayet star.
\par
The numbers of stars on the main sequence between 8 solar masses and about 25
solar masses, and between 25 solar masses and the upper end of the main
sequence also ought to correspond to the ratio of Wolf Rayet stars and the rest
of those stars which explode as supernovae.  Using for the simple estimate the
Salpeter mass function gives here an estimated ratio of about 1 in 5
supernova events which originate from a Wolf Rayet star.
\par
The model for the origin of the high energy population of relativistic cosmic
rays proposed in paper CR I and tested successfully in paper CR IV suggests
from
the energetics a ratio of about 1 in 3, assuming the amount of energy pumped
into
cosmic rays per supernova to be the same for all kinds. However, here we lump
all supernova explosions into stellar winds together, both with fast and slow
winds.
\par
Hence in the sample of radio supernovae of type II (6 sources)
presented by Weiler et al. (1986) there ought to be between none and two events
based on Wolf Rayet star explosions; in the sample of radio sources likely also
to be very young supernova remnants (28 sources with radio luminosities at 5
GHz) in the starburst galaxy M82 (Kronberg et al. 1985) there ought to be about
at least between 3 and 10 sources which originate from a Wolf Rayet star
explosion.  On the other hand, all of these objects are possibly explosions of
stars into former stellar winds (there is evidence that the initial mass
function in starburst galaxies is biased in favor of massive stars), and so the
proportion of the stars among them that are due to Wolf Rayet star explosions,
could be even higher than estimated here.  The radio luminosities are very
similar for the sources in M82 and the radiosupernovae of Weiler et al. (1986);
the average luminosity calculated in a variety of ways is always in the range
of
$3 \;10^{25} \, \rm erg \, sec^{-1} \, Hz^{-1}$ and $10^{26} \, \rm erg \,
sec^{-1} \, Hz^{-1}$, fitting our expectation (see above) for $U_1 \simeq 0.03
\,c$ rather well.   \par For our arguments on cosmic rays it is not relevant
whether the predecessor stars were Wolf Rayet stars or other massive stars with
extended winds permeated by strong magnetic fields.  But we conclude that a
sufficient fraction of the young radio supernova events and remnants are likely
to be from Wolf Rayet
stars, that we can use the implied properties for Wolf Rayet stars.

\titlea {Wolf Rayet stars}

The theoretical luminosities derived can be compared with the observations of
Wolf Rayet stars, which yield (Abbott et al. 1986) nonthermal luminosities up
to
about $5 \;10^{19} \;
\rm erg /sec /Hz$ at $5$ GHz and thermal luminosities up
to

$7 \;10^{18} \;\rm erg /sec /Hz$.  Since
the most important parameter, that enters here is the density of the wind, or
in
terms of wind parameters, the ratio of the mass loss to the wind speed $\dot{M}
/V_W$, we induce that the most extreme stars have a higher value for $\dot{M}
/V_W$ by $10.6$.  This translates into a higher nonthermal emission as well, by
a factor of $3.3$ to give  $6.0\,10^{24}\, {\rm erg/sec/Hz} \,
\eta_{-1}\, U_{1,-2}^2 \; B_{0.5}^{1.735}$, using the
case when free-free absorption dominates (for synchrotron self absorption the
corresponding luminosity is very nearly the same).  This is very much more than
the observed nonthermal luminosities and suggests that there is a limiting
factor.  We implicitly assume here, that some of the massive stars that
exploded
as the observed supernovae, either were Wolf Rayet stars before their
explosion,
or that their properties were not significantly different; since we derived the
wind density from the timing of the lightcurve above, and the shock velocity
both from the models of Owocki et al. (1988) and the argument that the shock
velocity be larger than the Alfv\'en velocity, which in turn we argued is not
very much lower than the wind velocity, the only other important parameter is
the magnetic field strength, and that we assume then to be of similar
magnitude.
Since Wolf Rayet stars do not distinguish themselves from stars that somewhat
later in life explode as supernovae, we can use all the same parameters, except
for the shock speed.  Then the only parameter left is the highly uncertain
efficiency of electron injection $\eta$.
\par
We can thus ask whether the efficiency $\eta$ might be
dependent on shock speed.  The shock speeds in supernovae are of order $0.03
\,c$ or larger, while those in Wolf Rayet star winds are of order $0.01 \,c$ or
less. Somewhere between these speeds there appears to be a critical shock
speed,
at which the character of electron injection changes.  One possible choice is
the following:  For shocks speeds near to $c \,(m_e/m_p)^{1/2}$
downstream thermalized electrons reach velocities close to the speed of light,
and so we might speculate that at shock speeds below and above this critical
velocity the injection process changes.  Here we do not wish to discuss
injection, but merely note that the data suggest efficiencies of order $\eta \,
\simeq \,10^{-6 \pm 1}$ for the shock speeds thought to be normal in Wolf Rayet
star winds, and of order $\eta \,
\simeq \, 0.1$ for supernova explosions.
The
data thus suggest that the injection of electrons into the diffusive shock
acceleration appears to exhibit a step function property at a shock velocity
near to $7\,10^8 \rm cm \,sec^{-1}$.  If true, this might be important also for
electron injection in quasars, which also exhibit a dichotomy into radioloud
and
radioweak objects.  In paper CR III (Biermann and Strom, 1993) we will then
test
whether this concept leads to a successful estimate for the electron/proton
ratio of the lower energy cosmic rays.
\par
Now we can go back and ask again, whether free-free absorption or synchrotron
self absorption dominates in the various cases considered:  Clearly, when
$\eta$, the efficiency for electron injection, is very small, then free-free
absorption always dominates, and so the maximum luminosities during a radio
variability episode of a Wolf rayet (or OB) star should be frequency dependent.
Also, for slow winds in supernova predecessor stars, again free-free absorption
will dominate normally (e.g. in the two examples used above).  On the other
hand, for fast winds of the supernova predecessor stars (as in the case of a
Wolf
Rayet star exploding), synchrotron self absorption is likely to be stronger
than
free-free absorption and then the maximum luminosities at different radio
frequencies should be independent of frequency.  This is a testable prediction,
since it relates radio and optical properties of a young supernova.

\titlea {OB stars}

The theoretical luminosities derived can be compared with the observations of
OB
stars, which yield (Bieging et al. 1989) nonthermal luminosities  up to about
$7 \;10^{19} \;
\rm erg /sec /Hz$ at $5$ GHz, and thermal luminosities up to
$2.5
\;10^{19} \;\rm erg /sec /Hz$.  Since the most important parameter, that enters
here is the density of the wind, or in terms of wind parameters, the ratio of
the mass loss to the wind speed $\dot{M} /V_W$, we induce that the most extreme
stars have a higher value for $\dot{M} /V_W$ by $27.6$.  This translates into a
higher nonthermal emission as well, by a factor of $5.4$ to give
$9.8\,10^{24}\,
{\rm erg/sec/Hz} \, \eta_{-1}\, U_{1,-2}^2 \; B_{0.5}^{1.735}$.  This is very
much more than the observed nonthermal luminosities.  Since the shock
properties
are likely to be similar to Wolf Rayet stars, this is again consistent with the
idea that the injection efficiency of electrons might be quite low, in
conjunction with magnetic field values not too far from what we argued to be
valid for Wolf Rayet stars.
\par
We can also compare the detection statistics and observed spectral indices:
Since in a variability episode a shock comes through from below through the
region of optical thickness near unity, the nonthermal emission increases
rapidly to its maximum, when its spectral index is

$$\alpha(\nu, max) \;=\; - 0.3 \, \alpha_{thin} \, ,\eqno\autnum $$
which is approximately $-0.2$; Thereafter, as the luminosity more slowly
decreases, the spectral index gradually approaches the optically thin index of
$-0.67$ or slightly steeper.  Therefore we expect the spectral index
distribution of detected sources to be a broad distribution from near $-0.2$ to
near $-0.7$; this is what has been found (Bieging et al. 1989).  Because of the
two possible signs of the magnetic field orientation, and the associated drift
energy gains for particles, we also expect that at any given time at most half
of all sources are detectable with nonthermal emission, even at extreme
sensitivity.  This is consistent with the observations.  We predict a similar
behaviour for Wolf Rayet stars.
\par
Here we have to ask how the magnetic fields can penetrate the radiative region
from below.  This question cannot presently be tackled by simulations on a very
large computer yet, but it is likely that the circulations induced by rotation
transport magnetic fields to the surface, where even a rather slight
differential rotation draws out the magnetic field into a mostly tangential
configuration.  In this case, clearly the origin of the momentum of the wind
can be readily accounted for from line driving (Lucy and Solomon 1970, Castor
et
al. 1975), and so we suspect that the magnetic driving adds only little (which
is the "other" case discussed above near the end of the wind section, section
8).
\par
We can also compare with the only existing theory to explain the nonthermal
radio emission of single massive stars with winds, by White (1985):  White's
theory is based on a concept involving a large number of shocks and does not
explain the data as already demonstrated by Bieging et al. (1989) in terms of
i) radio spectral index, ii) time variability nor iii) of the statistics of
detection.  Our theory is based on using single shocks and readily provides
spectral indices in the entire range that is observed, it explains the time
variability and easily accounts for a fair fraction of undetected sources.  On
the other hand, a second conclusion of Chen and White (1991b) and White and
Chen
(1992) is likely to be correct also in our theory, that OB and WR stars might
be
strong sources in the Gamma ray range from the secondaries resulting from
hadronic interactions of the energetic nuclei.

\titlea {An observational check}

Consider then the case where we observe the time evolution of the radio
emission
of massive stars:  We measure at maximum luminosity, and at two times later in
the optically thin regime the nonthermal luminosity again.  Then the ratio of
the two luminosities in the optically thin regime yields the zero point of the
shock event time.  The ratio of the luminosities at maximum and at one
of the optically thin epochs yields the time scale of expansion, the scaling
for
the shock event.  From an observation of the free-free emission, say, before
the
nonthermal outburst, we obtain the density and with that the radial scale,
where
the optical thickness is unity to free-free absorption.  The ratio of this
length scale and the explosion timescale gives the sum of the wind speed and
the
shock speed.  Given an estimate for the wind speed from UV data then yields the
shock speed.  And, finally, given the density, the shock speed and the
nonthermal luminosity we can obtain an estimate for the magnetic field.  We
note
that the shock travelling through the region where we normally have optical
depth of order unity to free-free absorption also causes the thermal emission
to
decrease since the shock increases the temperature drastically.  Hence the
total radio emission should reflect the increase of the nonthermal emission and
subsequent decrease, while at the same time the thermal emission will decrease
and subsequently recover.  Doing this exercise at several frequencies then
yields a control on the entire procedure, and also checks on the spectral index
dependence on wind speed.  Therefore, an intensive observing campaign at the
VLA
should enable us to check on our model in some detail.  Together with more
detailed modelling of the radio emission from radio supernovae such results
would then strengthen or weaken the case for the concept we propose here for
the acceleration of Cosmic Ray particles in the energy range from $10^4$ GeV to
$3\,10^9$ GeV.

\titlea {Conclusions}

In this paper we wanted to check whether the suggestion from cosmic ray
arguments that the magnetic field strengths in the wind of early type stars
are quite high, leads to any contradictions.  We are able to draw several
conclusions:
\par
1)  The magnetic fields in massive stars can be generated by a dynamo in the
convective interiors.  We can only speculate how they penetrate the
radiative zones during the main sequence phase. In those massive stars that do
not explode as supernovae, but become white dwarfs instead, the strong
magnetic field can become observable in the white dwarf stage.
\par
2)  The strong magnetic fields can put the major amount of momentum into wind
and accelerate them to slightly above the local Alfv\'en velocity; critical to
the argument is that the magnetic fields start already on the surface of the
star to be nearly tangential.
\par
3)  Indeed, with the parameters as suggested we are able to understand the
properties of radio supernovae, as regards i) the optically thin radio
spectrum, ii) the maximum luminosity, and iii) the time dependence of the
emission both during the optically thick and thin phase.  We derive a lower
limit to the magnetic field strength very close to the value implied by
cosmic ray arguments.
\par
4)  Using the same arguments for Wolf Rayet stars implies that the electron
injection is a step function of the shock velocity, with the step possibly
connected to the shock velocity when the thermal electron speed behind the
shock reaches relativistic values.
\par
5)  Similarly, the nonthermal radio emission of OB stars can be interpreted.
\par
6)  In summary, the strength of the magnetic field
in Wolf Rayet stars as estimated from fast magnetic rotator theory,
finds a variety of supportive arguments, from cosmic rays (papers CR
I and Stanev et al. 1993, paper CR IV), from the radioemission of radio
supernovae, and the radioemission of Wolf Rayet stars.  This demonstrates that
the proposal, that cosmic ray particles can be accelerated to extremely high
energies in the ultimate explosions of Wolf Rayet stars and other massive stars
with extended winds, up to energies of $3 \, 10^{9} \rm GeV$, is fully
consistent with the available data on radio supernova and Wolf Rayet stars.

%arXiv: \ack {
Most of the work by PLB was carried out during a five month sabbatical in
1991 at Steward Observatory at the University of Arizona, Tucson.  PLB
wishes to thank Steward Observatory, its director, Dr. P.A. Strittmatter, and
all the local colleagues for their generous hospitality during this time and
during many other visits.  PLB also wishes to thank Drs. D. Arnett, J.H.
Bieging, D. Eichler, T.K. Gaisser, J.R. Jokipii, F. Krause, J. Liebert, W.H.
Matthaeus, H. Meyer, R. Protheroe, T. Stanev, and R.G. Strom for extensive
discussions of Cosmic Ray, Supernova remnants and massive star physics. We wish
to thank Drs. T.K. Gaisser and J.H. Bieging for comments on the manuscript.
High energy physics with PLB is supported by DFG Bi 191/6,7, 9 (Deutsche
Forschungsgemeinschaft), DARA grant 50 OR 9202 of the BMFT (Bundesministerium
f\"ur Forschung und Technologie) and the NATO travel grant (to T. Stanev and
PLB) CRG 910072.  JPC work on this project was supported by NASA grant NAGW
2210.
%arXiv: }

\begref
\ref Abbott, D.C., Bieging, J.H., Churchwell, E., Torres, A.V.: 1986
     Astrophys.J. 303, 239
\ref Altenhoff, W., Mezger, P.G., Wendker, H., Westerhout, G.:  1960 Publ.Obs.
     Bonn, No.59, p. 48
\ref Bieging, J.H., Abbott, D.C., Churchwell, E.B.: 1989 Astrophys.J. 340, 518
\ref Biermann, P.L.:  1992 in "Frontiers in Astrophysics", Eds. Silberberg, R.,
     Fazio, G., Cambridge Univ. Press (in press since 1990, expected
     publication date June 1993)
\ref Biermann, P.L.:  1993 Astron\&Astroph.  271, 649 (paper CR I)
\ref Biermann, P.L., Chini, R., Greybe-G\"otz, A., Haslam, G., Kreysa, E.,
     Mezger, P.G.: 1990 Astron.\& Astroph. Letters 227, L21
\ref Biermann, P.L., Strom, R.G.:  1993 Astron.\&Astroph. (in press, paper CR
     III)
\ref Cassinelli, J.P.:  1982 in "Wolf-Rayet stars:  Observation, Physics,
     Evolution", Eds. C.W.H. de Loore, A.J. Willis, Reidel, p. 173
\ref Cassinelli, J.P.:  1991 in "Wolf-Rayet Stars and Interrelations with
     Other Massive Stars in Galaxies", IAU Sympos. No. 143,
     Eds. K.A van der Hucht and B. Hidayat, Kluwer, Dordrecht, p. 289
\ref Cassinelli, J.P.:  1993 a chapter to appear in a book in preparation
     with H.J.G.L.M. Lamers.
\ref Castor, J.I., Abbott, D.C., Klein, R.I.:  1975 Astrophys.J. 195, 157
\ref Chen, W., White, R.L.:  1991a Astrophys.J. Letters 381, L63
\ref Chen, W., White, R.L.:  1991b Astrophys.J. 366, 512
\ref Chevalier, R.A.:  1982 Astrophys.J. 259, 302
\ref Cox, J.P., Giuli, R.T.:  1968 "Principles of stellar structure", Gordon
and
     Breach, New York, 2 vols.
\ref Drury, L.O'C: 1983 Rep.Progr.Phys. 46, 973
\ref Gilbert, A.D., Childress, S.:  1990 Phys.Rev. Letters 65, 2133
\ref Gilbert, A.D.:  1991 Nature 350, 483
\ref Hartmann, L., Cassinelli, J.P.:  1982 B.A.A.S 13,795
\ref Hartmann, L., MacGregor, K.B.:  1982 Astrophys.J. 259, 180
\ref Hidayat, B.:  1991 in "Wolf-Rayet Stars and Interrelations with
     Other Massive Stars in Galaxies", IAU Sympos. No. 143,
     Eds. K.A van der Hucht and B. Hidayat, Kluwer, Dordrecht, p. 587
\ref Jauncey, D.L., Kemball, A., Bartel, N., Whitney, A.R., Rogers, A.E.E.,
     Shapiro, I.I., Preston, R.A., Clark, T.A., Harvey, B.R., Jones, D.L.,
     Nicolson, G.D., Nothnagel, A., Phillips, R.B., Reynolds, J.E., Webber,
     J.C.:  1988 Nature 334, 412
\ref Jokipii, J.R., Morfill, G.:  1987 Astrophys.J. 312, 170
\ref Kato, M., Iben, I.Jr.:  1992 Astrophys.J. 394, 305
\ref Kirk, J.G., Wassmann, M.:  1991 Preprint
\ref Kronberg, P.P., Biermann, P.L., Schwab, F.R.:  1985 Astrophys. J. 291, 693
\ref Lagage,P.O., Cesarsky, C.J.: 1983 Astron.\& Astroph. 118, 223
\ref Langer, N.:  1989 Astron.\&Astroph. 210, 93
\ref Leitherer, C.:  1991 in "Wolf-Rayet Stars and Interrelations with
     Other Massive Stars in Galaxies", IAU Sympos. No. 143,
     Eds. K.A van der Hucht and B. Hidayat, Kluwer, Dordrecht, p. 465
\ref Lucy, L.B., Solomon, P.M.:  1970 Astrophys.J. 159, 879
\ref MacFarlane, J.J., Cassinelli, J.P.:  1989 Astrophys.J. 347, 1090
\ref Maheswaran, M., Cassinelli, J.P.:  1988 Astrophys.J. 335, 931
\ref Maheswaran, M., Cassinelli, J.P.:  1992 Astrophys.J. 386, 695
\ref Massey, P., Armandroff, T.E.:  1991 in "Wolf-Rayet Stars and
Interrelations
with
     Other Massive Stars in Galaxies", IAU Sympos. No. 143,
     Eds. K.A van der Hucht and B. Hidayat, Kluwer, Dordrecht, p. 575
\ref Mezger, P.G., Henderson, A.P.: 1967 Astrophys.J. 147, 471
\ref Nerney, S., Suess, S.T.:  1987 Astrophys.J. 321, 355
\ref Owocki, S.P., Castor, J.I., Rybicki, G.B.:  1988 Astrophys.J. 335, 914
\ref Panagia, N., Sramek, R.A., Weiler, K.W.:  1986 Astrophys.J. Letters 300,
     L55
\ref Parker, E.N.:  1958 Astrophys.J. 128, 664
\ref Poe, C.H., Friend, D.B., Cassinelli, J.P.:  1989 Astrophys.J. 337, 888
\ref Rachen, J.: 1992 Masters of Science Thesis, University of Bonn
\ref Rachen, J., Biermann, P.L.:  1992 in Proc. "Particle acceleration in
cosmic
     plasmas", Eds. G.P. Zank, T.K. Gaisser, AIP conf. No. 264, p. 393
\ref Rachen, J., Biermann, P.L.:  1993 Astron.\&Astroph. 272, 161,
     (paper UHE  CR I)
\ref Rachen, J., Stanev, T., Biermann, P.L.:  1993 Astron.\&Astroph.
     (in press, paper UHE CR II)
\ref Rybicki, G.B., Lightman, A.P.:  1979 Radiative Processes in Astrophysics,
     Wiley Interscience, New York
\ref Ruzmaikin, A., Sokoloff, D., Shukurov, A.:  1988 Nature 336, 341
\ref Schmidt, G.D., Bergeron, P., Liebert, J., Saffer, R.A.:  1992 Astrophys.J.
     394, 603
\ref Schmutzler, T.:  1987 Ph.D. thesis, University of Bonn
\ref Shara, M.M., Potter, M., Moffat, A.F.J., Smith, L.F.:  1991 in
    "Wolf-Rayet Stars and Interrelations with
     Other Massive Stars in Galaxies", IAU Sympos. No. 143,
     Eds. K.A van der Hucht and B. Hidayat, Kluwer, Dordrecht, p. 591
\ref Silberberg, R., Tsao, C.H., Shapiro, M.M., Biermann, P.L.: 1990
     Astrophys.J. 363, 265
\ref Stanev, T., Biermann, P.L., Gaisser, T.K.:  1993 Astron. \&Astroph.
     (in press, paper CR IV)
\ref Staveley-Smith, L., Manchester, R.N., Kesteven, M.J., Campbell - Wilson,
     D., Crawford, D.F., Turtle, A.J., Reynolds, J.E., Tzioumis, Killeen,
N.E.B.,
     Jauncey, D.L.:  1992 Nature 355, 147
\ref Turtle, A.J., Campbell-Wilson, D., Bunton, J.D., Jauncey,
     D.L., Kesteven, M.J., Manchester, R.N., Norris, R.P., Storey, M.C.,
     Reynolds, J.E.:  1987 Nature 327, 38
\ref V\"olk,H.J., Biermann, P.L.: 1988 Astrophys.J. Letters 333, L65
\ref Weber, E.J., Davis, L.Jr.:  1967 Astrophys.J. 148, 217
\ref Weiler, K.W., Sramek, R.A., Panagia, N., Hulst, J.M. van der, Salvati, M.:
     1986 Astrophys.J. 301, 790
\ref Weiler, K.W., Panagia, N., Sramek, R.A., Hulst, J.M. van der, Roberts,
     M.S., Nguyen, L.:  1989 Astrophys.J. 336, 421
\ref Weiler, K.W., Panagia, N., Sramek, R.A.:  1990 Astrophys.J. 364, 611
\ref Weiler, K.W., Dyk, S.D. van, Panagia, N., Sramek, R.A., Discenna, J.L.:
     1991 Astrophys.J. 380, 161
\ref White, R.L.:  1985 Astrophys.J. 289, 698
\ref White, R.L., Long, K.S.:  1986 Astrophys.J. 310, 832
\ref White, R.L., Chen, W.:  1992 Astrophys.J. Letters 387, L81
\endref
\bye